\documentclass[usegraphicx,useAMS,usenatbib]{mn2e}

\usepackage{amssymb}
\usepackage{times}
\usepackage{aas_macros} %needed to define journal macros generated by
\title[Multisite campaign on M67. II.]{Multisite campaign on the open
  cluster M67. II. Evidence for solar-like oscillations in red giant stars} 
\author[D. Stello et al.]
{D. Stello$^{1,2,3}$,
H. Bruntt$^{1,2,3}$,
H. Kjeldsen$^{1,4}$,
T. R. Bedding$^{2}$,
T. Arentoft$^{1,2,4}$,\newauthor
R. L. Gilliland$^{5}$,
J. Nuspl$^{6}$,
S.-L. Kim$^{7}$,
Y. B. Kang$^{7}$,
J.-R.~Koo$^{7}$,
J.-A.~Lee$^{7}$,\newauthor
C. Sterken$^{8}$,
C.-U. Lee$^{7}$,
H. R. Jensen$^{1}$,
A. P. Jacob$^{2}$,
R.~Szab\'o$^{9,6}$,
S.~Frandsen$^{1,4}$,\newauthor
Z. Csubry$^{6}$,
Z.~E. Dind$^{2}$,
M. Y. Bouzid$^{8}$,
T.~H. Dall$^{10}$ and
L.~L.~Kiss$^{2}$\\
$^{1}$Institut for Fysik og Astronomi (IFA), Aarhus Universitet, 8000 Aarhus,
Denmark\\
$^{2}$School of Physics, University of Sydney, NSW 2006, Australia\\
$^{3}$Department of Physics, US Air Force Academy, Colorado Springs, CO
80840, USA\\
$^{4}$Danish AsteroSeismology Centre, Aarhus Universitet, DK-8000 Aarhus,
Denmark\\
$^{5}$Space Telescope Science Institute, 3700 San Martin Dr., Baltimore, USA\\
$^{6}$Konkoly Observatory of the Hungarian Academy of Sciences, H-1525 Budapest, PO Box 67, Hungary\\
$^{7}$Korea Astronomy and Space Science Institute, Daejeon 305-348, Korea\\
$^{8}$Vrije Universiteit Brussel, Pleinlaan 2, B-1050 Brussels, Belgium\\
$^{9}$Physics Department, University of Florida, Gainesville, FL, 32611, USA\\
$^{10}$European Southern Observatory, Casilla 19001, Santiago 19, Chile}

\begin{document}

\date{Accepted 2006 December 15. Received 2006 December 14; in original
  form 2006 October 11} 

\pagerange{\pageref{firstpage}--\pageref{lastpage}} \pubyear{2006}

\maketitle

\label{firstpage}

\begin{abstract}
%{\bf Context:} 
Measuring solar-like oscillations in an ensemble of stars 
in a cluster, holds promise for testing stellar structure and
evolution more stringently than just fitting parameters to single
field stars. The most ambitious attempt to pursue these prospects was by  
\citet{Gilliland93} who targeted 11 turn-off stars in the open cluster M67
(NGC 2682), but the oscillation amplitudes were too small ($<20\,\mu$mag)
to obtain unambiguous detections. 
%{\bf Aims:}
Like \citet{Gilliland93} we also aim at detecting solar-like
oscillations in M67, but we target red giant stars with expected
amplitudes in the range 50--500$\,\mu$mag and periods of 1 to 8 hours.
%{\bf Methods:} 
We analyse our recently published photometry measurements, obtained during a
six-week multisite campaign using nine telescopes around the world.
The observations are compared with simulations and with
estimated properties of the stellar oscillations. 
%{\bf Results:} 
Noise levels in the Fourier spectra as low as $27\,\mu\rmn{mag}$ are obtained
for single sites, while the combined data reach $19\,\mu\rmn{mag}$, making
this the best photometric time series of an ensemble of red giant
stars. These data enable us to make the first test of the scaling
relations (used to estimate frequency and amplitude) with an homogeneous
ensemble of stars. 
%We see a systematic shift in frequency of excess power as a function of
%luminosity when the stars are analysed as an ensemble
%{\bf Conclusion:} 
The detected excess power is consistent with the expected signal from
stellar oscillations, both in terms of its frequency range and
amplitude. However, our results are limited by apparent high levels of 
non-white noise, which cannot be clearly separated from the stellar
signal.   
\end{abstract}

\begin{keywords}
stars: red giants -- stars: oscillations -- stars variables: other -- 
open clusters and associations: individual: M67 (NGC 2682) -- Techniques:
photometric. 
\end{keywords}

% {\it IRAS\/} <= LOOK AT THAT for italic

%use \hbox{60\,$\umu$m} <= LOOK AT THAT \hbox for no-brake and \umu for
%non-italic greek 

%\section[]{Decription of the Envelope\\* Model}(Envelope\\* Model) <=
%LOOK AT THAT 

%\begin{equation}
%   L(\nu)=\mskip-12mu\int\limits_{\rmn{envelope}}
%   \mskip-12mu (\mskip-12mu) <= LOOK AT THAT
%   \rho(r)Q_{\rmn{abs}}(\nu)B[\nu,T_{\rmn{g}}(r)]
%   \exp [-\tau(\nu,r)]\>  (\>) <= LOOK AT THAT I think its a space marker
%\end{equation}

%${\balpha}$ for boldface greek

%-----------------------------------------------------------------------
\section{Introduction}
The first clear detection of solar-like oscillations in a red giant star
\citep[$\xi\,\rmn{Hya}$;][]{Frandsen02,Stello02}
opened up a new part of the Hertzsprung-Russell diagram to be explored with
asteroseismic techniques. 
Following that discovery, detailed analysis of $\xi\,\rmn{Hya}$ has been
performed \citep{Teixeira03,Stello04,Stello06} and new
discoveries of oscillations in similar stars have emerged
\citep[$\varepsilon\,\rmn{Oph}$ and $\eta\,\rmn{Ser}$;][]{Barban04,Ridder06}.
These results are all based on radial velocity
measurements of high precision ($\sigma\sim2\,$m/s) but from non-continuous 
observations, which imposes large ambiguities on the results
\citep{Stello06,Ridder06}.  
With oscillation periods of a few hours,
these stars require a time base of roughly one month, which can only be
obtained on small telescopes. But the current lack of high-precision
spectrographs %(similar to CORALIE and ELODIE \citet{ref}) 
on small telescopes makes a multisite campaign impossible.

\begin{figure}
 \includegraphics[width=88mm]{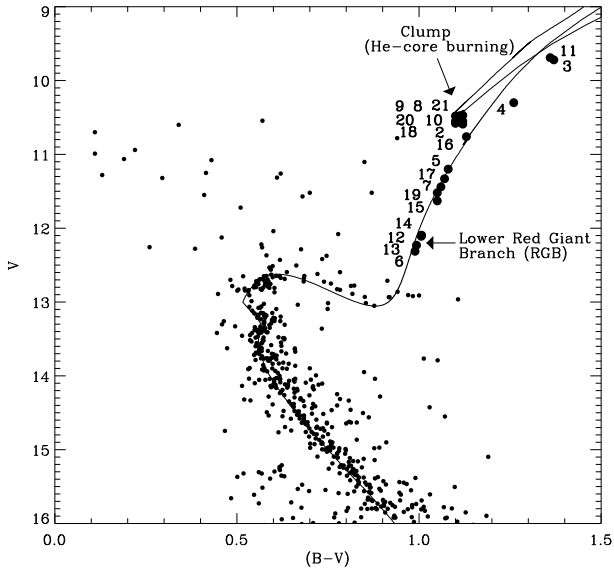}
 \caption{Colour-magnitude diagram of the open cluster M67 
 \citep[photometry from][]{Montgomery93}. The target stars are indicated with identifier numbers
 corresponding to those given in Tables~\ref{tab2} and \ref{tab4}. The solid
 line is an isochrone with $(m-M)=9.7\,\rmn{mag}$, $\rmn{Age}=4.0\,\rmn{Gyr}$,
 $Z=0.0198$ and $Y=0.2734$ from the BaSTI database \citep{Pietrinferni04}.}  
\label{fig0}
\end{figure}

However, using photometry makes it feasible to incorporate many 1--2m
class telescopes in a multisite campaign, and it furthermore provides the  
possibility to observe many stars, like in a cluster, simultaneously. 
Detecting oscillations in a set of cluster stars potentially increases
the power of the asteroseismic 
measurements due to the additional constraints provided by the common
parameters 
of the cluster members (age and composition).
Until recently, there has been no such ground-based photometric campaign
aimed at detecting solar-like oscillations in red giant stars.
A number of attempts have been made to detect oscillations in more Sun-like
stars (hotter and less luminous) of the open cluster M67 
\citep{GillilandBrown88,Gilliland91,GillilandBrown92b}, with the
most ambitious multisite effort made by
\citet{Gilliland93}. Despite noise levels as low as $0.29\,\rmn{mmag}$
per minute integration no unambiguous detections were claimed.
In a recent paper \citep[][~hereafter Paper I]{Stello06a}, we reported
observations from a large six-week multisite campaign also aimed at M67. 
However, unlike the previous studies our campaign was 
optimized for the slightly brighter and longer period red giant stars (see
Fig.~\ref{fig0}), and also covered a much longer time span. The long
oscillation periods mean that non-white noise such as drift is more
crucial for this project than in the previous studies. The
data set reported in Paper I was based on very different sites, many with
unknown long-term stability performances. A realistic estimate of the
final non-white noise in the data could therefore not be obtained prior to
observations.  
%We have one target star in common with \citet{Gilliland93}
%one in common with \citet{GillilandBrown92b} and seven targets are also in
%\citet{Gilliland91}. 

In this paper our main emphasis is on the time series analysis of the red
giant stars (Sect.~\ref{analysis}), based on the data described in Paper
I. We report in Sect.~\ref{observations} on an additional independent data
reduction method to further obtain lower noise in the Fourier spectra. In
Sect.~\ref{expectsignal} we estimate the oscillation characteristics and
simulate in Sect.~\ref{simulations} the expected outcome for each target
without the presence of non-white noise to facilitate the analysis of the
observations. We give our conclusions in Sect.~\ref{conclusions}.

%-----------------------------------------------------------------------
\section[]{Observations and data reduction}\label{observations}
The data are from a global multisite observing campaign with nine 0.6-m to
2.1-m class telescopes from 6 January to 17 February 2004
(Paper I). The photometric time series of those stars within the
field-of-view of all telescopes comprises roughly 18000 data points. 

After calibrating the CCD images we used the {\sevensize MOMF} 
package \citep{KjeldsenFrandsen92}, which 
calculated differential time series of 20 red giants relative
to a large ensemble of 
stars (from 116 to 358 stars, depending on telescope field-of-view).
We performed the following three initial steps to improve the signal-to-noise
in the final Fourier spectra of the time series: (1) sigma clipping, (2)
correction for colour extinction and (3) calculation of weights for each data
point. For further details see 
Paper I. %Evt lidt mere om data (fra Paper I) hvis relevant.

In addition to the time series produced by D.S., as described above,
H.B. constructed differential time series following the approach of 
\citet{Honeycutt92}. This was based on the raw time series (also calculated
by {\sevensize MOMF}), but using only the red giant stars as the
ensemble. 
For each target star, the differential time series was calculated by
subtracting a reference time series that did not include the star
itself. %IS THAT RIGHT  
The reference time series comprised offsets for each data point (CCD
image), which were a weighted average of the ensemble.
For each data point the weight $w$ given to each star was
calculated as $w=1/(\sigma_{\rmn{ptp}}+\sigma_{\rmn{min}})$, where
$\sigma_{\rmn{ptp}}$ is the local point-to-point scatter and
$\sigma_{\rmn{min}}=1\,$mmag is a fixed minimum noise value to prevent a
single star with very low noise from dominating the ensemble.  
Using the relatively small homogeneous 
ensemble comprising only red giants has the advantage of better removing
colour-dependent extinction, and hence providing a lower noise
level in the Fourier spectrum at low frequencies. However, simulations
showed that the reference time series will include stellar oscillations 
with amplitudes up to 30\% of those we want
to detect. The stars with the longest periods (hence largest amplitudes)
will be the most affected because the sample is dominated by these.
%HB what are the weights for each star when you calc the reference series?
%evt fig der viser Sim Kitt: mean of all RG

In most stars, the two methods produced very similar noise levels 
in the time series, within 10\%. However, on a few
stars differences of up to 50\% were seen. 
In the following, we  
chose the time series for each star and each site with the lowest noise
in the Fourier spectrum in the interval 300--900$\,\mu$Hz, which is 
outside the frequency range where the stars are expected to
oscillate. 
%Although previous investigations by 
%\citet{Gilliland93} showed a gain of a few percent in the average time
%series based on different reduction methods, we see no significant gain
%within the frequency range of the expected stellar oscillations in the
%average time series from the two reductions of D.S. and H.B. 

\begin{table*}
 \centering
 \begin{minipage}{140mm}
  \caption{Properties of red giant target stars.}
  \begin{tabular}{@{}rrrrrrr@{}lrrr@{/}lr@{}}
  \hline
   No & $V$\footnote{Photometry from \citet{Montgomery93}.} & $B-V$ &
   $L/\rmn{L}_{\sun}$ & $T_{\rmn{eff}}$ &   
   \multicolumn{3}{c}{$\delta L/L$  $\mu$mag\footnote{Estimated amplitudes
   based on both $(L/M)$ and $(L/M)^{0.7}$ scaling. Note that
   $1\,\rmn{ppm}=1.086\,\mu$mag.}} & 
   $\nu_{\rmn{max}}$ & $\Delta\nu_{0}$ & 
   \multicolumn{2}{c}{Cross-ref\footnote{IDs starting with an S are from
   \citet{Sanders77}, while G are from \citet{Gilliland91}.}} &
   P\footnote{Membership probabilities are from \citet{Zhao93} and
   \citet{Sanders77} respectively. All targets have high probabilities in
   \citet{Girard89} ($\rmn{P}>95\%$).}\\  
      & mag &  mag  &              &               K &   
   \multicolumn{1}{c}{$(L/M)$} & \multicolumn{2}{r}{$(L/M)^{0.7}$} &
   $\mu$Hz        &    $\mu$Hz & \multicolumn{2}{c}{}  &  \%\\ 
 \hline
 3&  9.72& 1.37& 250.6& 3920& 2080& ~~434& &   3.7&  0.7&   S978&G8 & 53/95 \\
11&  9.69& 1.36& 243.5& 3960& 1980& ~~417& &   3.9&  0.8&  S1250&G4 & 58/95 \\
 4& 10.30& 1.26&  87.0& 4330&  592& ~~170& &  15.1&  2.2&  S1016&G5 & 78/93 \\
21& 10.47& 1.12&  51.9& 4727&  296& ~~ 99& &  34.4&  4.2&  S1592&   & 48/92 \\
 8& 10.48& 1.11&  50.8& 4750&  287& ~~ 97& &  35.8&  4.3&  S1010&G2 & 82/96 \\
 9& 10.48& 1.10&  50.2& 4772&  281& ~~ 95& &  36.8&  4.4&  S1084&   & 80/92 \\
10& 10.55& 1.12&  48.2& 4727&  275& ~~ 94& &  37.0&  4.4&  S1279&G7 & 79/92 \\
20& 10.55& 1.10&  47.1& 4772&  264& ~~ 91& &  39.2&  4.6&  S1479&   & 76/95 \\
 2& 10.59& 1.12&  46.4& 4727&  265& ~~ 92& &  38.4&  4.6&  S1074&   & 74/91 \\
18& 10.58& 1.10&  45.8& 4772&  256& ~~ 89& &  40.3&  4.7&  S1316&   & 73/95 \\
16& 10.76& 1.13&  40.3& 4703&  232& ~~ 84& &  43.5&  5.0&  S1221&   & 92/90 \\
 5& 11.20& 1.08&  25.4& 4815&  140& ~~ 58& &  74.8&  7.6&  S1054&G9 & 93/64 \\
17& 11.33& 1.07&  22.4& 4835&  122& ~~ 53& &  86.0&  8.4&  S1288&   & 94/96 \\
 7& 11.44& 1.06&  20.2& 4854&  109& ~~ 49& &  96.9&  9.2&   S989&G12& 95/95 \\
19& 11.52& 1.05&  18.7& 4873&  101& ~~ 46& & 106.0&  9.9&  S1254&   & 94/95 \\
15& 11.63& 1.05&  16.9& 4873&   91& ~~ 43& & 117.3& 10.6&  S1277&   & 95/95 \\
14& 12.09& 1.01&  11.2& 4945&   58& ~~ 31& & 187.3& 15.2&  S1293&   & 96/93 \\
12& 12.11& 1.01&  11.0& 4947&   57& ~~ 31& & 190.2& 15.4& S1264a&G15& 0/75 \\
13& 12.23& 0.99&   9.9& 4966&   51& ~~ 28& & 213.2& 16.8&  S1305&   & 96/95 \\
 6& 12.31& 0.99&   9.2& 4971&   48& ~~ 27& & 230.2& 17.8&  S1103&   & 95/---\\
%1103 is star MMJ5663/Zhao578 Zhao is:A&AS100,243
\hline
\vspace{-1cm}
\label{tab2}
\end{tabular}
\end{minipage}
\end{table*}

%-----------------------------------------------------------------------
\section{Expected oscillation signal}\label{expectsignal}
To better interpret our results, we have estimated the
characteristics of solar-like oscillations expected in the red giant stars.
We scaled the oscillation parameters of the Sun to predict amplitude,
central frequency of excess power and the 
large frequency separation. These predictions are used in
Sects.~\ref{analysis} and \ref{simulations} to compare with the
observations. 

The predicted amplitude in the Johnson $V$ filter
($\lambda_\rmn{cen}=544\,\rmn{nm}$)  
was derived using the scaling relation by \citet{KjeldsenBedding95} (using
$1\,\rmn{ppm}=1.086\,\mu$mag): 
\begin{equation}
 (\delta L/L)_{\lambda} = 
             {L/\rmn{L}_{\sun}(5.1\pm 0.3)\,\mu\rmn{mag} \over 
              (\lambda/550\,\rmn{nm})(T_{\rmn{eff}}/5777\,\rmn{K})^2(M/M_{\sun}) }\,.
\label{loverm}
\end{equation}
These amplitudes were used as a guide while planing the observations. 
However, recent theoretical studies indicate that the $L/M$-scaling may
over-estimate the amplitude 
%\citep{Stello02,Ridder06} 
for main-sequence stars and that $(L/M)^{0.7}$-scaling might provide a more 
realistic prediction 
\citep{Samadi05}, which we will take into account in evaluating of our
results. We note that extrapolating amplitudes for red giant stars based on
these scaling relations is very uncertain, and has so far not
been thoroughly tested by calculations of theoretical pulsation models
or observations of these stars.

The characteristic frequency domain within which a star is
oscillating was estimated as the central frequency of the excess power,
which was obtained by scaling the acoustic cut-off frequency of the Sun
\citep{Brown91} 
\begin{eqnarray}
 \nu_{\rmn{max}} 
    &=&  \frac{M/\rmn{M}_{\sun}} {(R/\rmn{R}_{\sun})^2\,
         \sqrt{T_{\rmn{eff}}/5777\,\rmn{K}}}\times 3050\,\mu\rmn{Hz}\nonumber\\
    &=&  \frac{M/\rmn{M}_{\sun}\,(T_{\rmn{eff}}/5777\,\rmn{K})^{3.5}} 
         {L/\rmn{L}_{\sun}}\times 3050\,\mu\rmn{Hz}\, ,
\label{eq_freq_max}
\end{eqnarray}
\noindent 
using $L/\rmn{L}_{\sun}=(R/\rmn{R}_{\sun})^2(T_{\rmn{eff}}/5777\,\rmn{K})^4$. 
This scaling relation gives very good agreement with the frequency range
of the solar-like oscillations observed in main sequence stars and also in
red giants \citep{BeddingKjeldsen03}. 

Finally, we predict the expected frequency spacing, $\Delta\nu_{0}$,
between modes of the same degree in the power spectrum
\citep{KjeldsenBedding95} 
\begin{eqnarray}
\Delta\nu_{0} 
    &=& \frac{ (M/\rmn{M_{\sun}})^{0.5} } {(R/\rmn{R_{\sun}})^{1.5}}
              \times (134.92\pm 0.02)\,\mu\rmn{Hz}\\
    &=& \frac{ (M/\rmn{M_{\sun}})^{0.5} (T_{\rmn{eff}}/5777\,\rmn{K})^3}
             {(L/\rmn{L}_{\sun})^{0.75}} 
              \times (134.92\pm 0.02)\,\mu\rmn{Hz}\ .\nonumber
\label{eq_large_sep}
\end{eqnarray}

To calculate these parameters we made rough estimates of the
stellar mass, luminosity, and effective temperature, which are summarized in
Table~\ref{tab2}. 
For all stars we adopted a mass of $M=1.35\rmn{M}_{\sun}$ corresponding to the
mass at the base of the red giant branch for an isochrone with
$\rmn{Age}=4.0\,\rmn{Gyr}$, $Z=0.0198$ and $Y=0.2734$ 
\citep[BaSTI database;][]{Pietrinferni04}, which matches the cluster
colour-magnitude diagram (see Fig.~\ref{fig0}). 
This is also in good agreement with the turn-off mass by 
\citet{VandenBergStetson04}.
%error max 10% in mass
We derived $L$ and $T_{\rmn{eff}}$ using ($V$, $B-V$)-photometry
of the cluster \citep{Montgomery93} and interpolation of the BaSeL grid
\citep{Lejeune98}. 
For our purpose it was sufficient to adopt a typical surface gravity for
all red giants of log$\,g$=2.5 \citep{Allen73}, [Fe/H]=0.0 in agreement
with \citet{Nissen87}, and $(m-M)=9.7\,\rmn{mag}$ corresponding to
$d=870\,$pc, which is within 5\% of 
previous investigations \citep[e.g.][]{Montgomery93,VandenBergStetson04}. 
%Our derived $L$ and  $T_{\rmn{eff}}$ values are in excellent
%agreement with the isochrone of \citet{Pietrinferni04} 
%(see Fig. 1 in Paper I). 
The assumption of a common log$\,g$ affected our tempareture estimates
by less than 100$\,$K and our luminosity estimates by up to 10\% for the
stars investigated.
We note that, because the temperatures of our target stars are
roughly the same, the expected amplitudes are proportional to
$1/\nu_{\rmn{max}}$ 
\begin{equation}
  \left(\frac{\delta L}{L}\right)_{\lambda}
        = \frac{1}{\nu_{\mathrm{max}}}
          \left(\frac{T_{\mathrm{eff}}}
%                     {\mathrm{T}_{\mathrm{eff},\odot}}\right)^{1.5} 
                     {5777\,\rmn{K}}\right)^{1.5} 
          \frac{550\,\rmn{nm}}{\lambda} (10.1\pm 0.8)\,\mathrm{mmag}. 
\label{eq_l_stello_scale}
\end{equation}
\noindent 

\begin{table*}
 \centering
 \begin{minipage}{160mm}
  \caption{Mean noise level (in $\mu\rmn{mag}$) in the Fourier spectrum for two
  frequency intervals (stars plotted in Fig.~\ref{fig1} are in
  boldface). Site abbreviations are:  
          SSO$_1$ (Wide Field Imager at Siding Spring Observatory,
          Australia); 
          SSO$_2$ (Imager at Siding Spring); 
          SOAO (Sobaeksan Optical Astronomy Observatory, Korea);  
          SAAO (South Africa Astronomical Observatory); 
          RCC (Ritchey-Chr\'etien-Coud\'e at Piszk\'estet\H{o}, Konkoly
          Observatory, Hungary);  
          Sch (Schmidt at Piszk\'estet\H{o}); 
          LaS (La Silla Observatory, Chile); 
          LOAO (Mt. Lemmon Optical Astronomy Observatory, Arizona);  
          Kitt (Kitt Peak National Observatory, Arizona); 
          Lag (Mt. Laguna Observatory, California).}
  \begin{tabular}{@{}rr@{, }rr@{, }rr@{, }rr@{, }rr@{, }rr@{, }rr@{, }rr@{, }rr@{, }rr@{, }rr@{, }rr@{, }rr@{, }rr@{, }rr@{, }rr@{, }rr@{, }rr@{, }rr@{, }rr@{, }r@{}}
  \hline
No & \multicolumn{2}{c}{SSO$_1$} & \multicolumn{2}{c}{SSO$_2$} &
\multicolumn{2}{c}{SOAO} & \multicolumn{2}{c}{SAAO} &
\multicolumn{2}{c}{RCC} & \multicolumn{2}{c}{Sch} & \multicolumn{2}{c}{LaS}
& \multicolumn{2}{c}{LOAO} & \multicolumn{2}{c}{Kitt} &
\multicolumn{2}{c}{Lag} \\ 
\multicolumn{20}{c}{$\sigma_{1000\rmn{-}3000\,\mu\rmn{Hz}},
  \sigma_{300\rmn{-}900\,\mu\rmn{Hz}}$ }\\ 
 \hline
%Nedenstaaende er stoejen den bedst af min og HBs fotometri for gver stj og site
 3& 291&372&  298&311&  400&535&   --& --&   --& --&  809&1090&  481&585&
 349&545&   --&--&  604&746\\ 
11&2930&3325& 231&235&  334&366&  294&417&   --& --& 1096&1737&  102&120&
 476&688&   --& --&  152&196\\ 
{\bf 4}&  99&110&  138&134&  147&137&   43& 48&  248&386&  139&144&   39& 41&
 164&268&   62&91&  123&173\\ 
21&  --& --&  229&260&  359&394&   --& --&   --& --&  227&302&   --& --&
 149&153&   --& --&   --& --\\ 
{\bf 8}& 101& 87&  155&155&  117&134&   48& 52&  171&295&  117&111&   37& 38&
 146&229&   47& 59&  118&152\\ 
{\bf  9}&  94& 95&  154&174&  129&141&   --& --&   --& --&  120&130&   40& 42&
 194&315&   27& 31&  118&157\\ 
{\bf 10}&  82& 69&   97&122&  119&115&   68& 78&  164&284&  113&124&   37& 40&
 147&217&   34& 42&   93&124\\ 
 20&  86& 83&  115&119&  152&157&   --& --&   --& --&  150&196&   39& 39&
 258&465&   --& --&  191&336\\ 
{\bf  2}&  96&101&  135&157&  148&151&   --& --&   --& --&  138&138&   53& 52&
 150&202&   32& 38&  125&166\\ 
{\bf 18}&  77& 78&  122&134&  187&176&   --& --&   --& --&  120&136&   33& 36&
 247&451&   --& --&  156&280\\ 
16&  92&106&  167&164&  259&239&   --& --&   --& --&  150&150&   46& 48&
 289&312&   --& --&  184&295\\ 
{\bf  5}&  96& 87&  160&171&  213&222&   45& 47&  219&324&  165&186&   57& 59&
 108&128&   43& 51&  114&175\\ 
{\bf 17}& 100&103&  126&148&  199&194&   --& --&  255&341&  162&168&  101&122&
 154&303&   36& 43&  163&225\\ 
{\bf  7}& 108& 95&  159&178&  250&235&   66& 81&  319&518&  171&170&   46& 50&
 122&160&   39& 47&  511&810\\ 
19& 113&103&  160&180&  219&202&   --& --&   --& --&  182&180&   55& 64&
 142&235&  138&230&  153&191\\ 
{\bf 15}& 109& 91&  153&145&  237&273&   --& --&  331&453&  175&225&  163&198&
 101&116&   45& 53&  131&184\\ 
{\bf 14}& 109& 94&  178&192&  330&340&   --& --&  414&563&  205&237&   50& 51&
 117&129&   63& 67&  131&185\\ 
12& 231&240&  486&567&  719&692&  140&156&  471&470&  198&227&  276&317&
 371&587&   --& --&  224&299\\ 
{\bf 13}& 111&104&  187&192&  349&336&   --& --&   --& --&  224&227&  111&128&
  96&119&   74& 85&  148&196\\ 
 6&  --& --&  231&260&  351&378&   --& --&   --& --&  232&289&   --& --&
 161&186&   --& --&   --& --\\ 
\hline
%\vspace{-1cm}
\label{tab4}
\end{tabular}
\end{minipage}
\end{table*}

%-----------------------------------------------------------------------
\section{Time series analysis}\label{analysis}

In Table~\ref{tab4} we give for each star the noise levels (in amplitude)
in the Fourier spectra measured in two frequency intervals based on
single-site data. The noise denoted $\sigma_{1000\rmn{-}3000\,\mu\rmn{Hz}}$
represent the lowest noise level (white noise), while
$\sigma_{300\rmn{-}900\,\mu\rmn{Hz}}$ is the noise level closer to, but
still outside the expected frequency range of the oscillations.
The best data were from SAAO, La Silla and Kitt Peak, with mean noise levels 
of roughly 40$\,\mu$mag for the best stars. However, at SAAO only a
few stars were observed due to the small field-of-view. The Kitt Peak data
generally had a slightly lower point-to-point scatter but were more
affected by extinction and hence showed more drift noise. 
As we seek to optimize the signal-to-noise in the
final Fourier spectra (in amplitude) based on the combined data from all
sites, the data from La Silla and Kitt Peak will dominate for the majority
of the stars. For further details about the noise properties see Paper I.

Due to nightly drifts in the data we saw very strong peaks at 1--4
cycles per day (corresponding to 11.57, 23.15, 34.72 and  46.30$\,\mu$Hz) 
in the Fourier spectra based on individual sites. Even when
combined, the data still showed significant excess power due to these
drifts. This was a serious problem because the most promising stars in the  
ensemble are expected to oscillate in the affected frequency range.
% at these frequecies. 
%Due to aliasing these peaks produced a dominating series of peaks in
%the spectra, which obscured the frequency analysis. 
%In addition we often saw a significant peak at 0.5 cycle per day which was
%due to a near singularity in the calculation of the spectrum
%\citep{Frandsen95}. 
We decided to remove (``clean'') these specific frequencies using standard 
iterative sine-wave fitting, on a site-by-site basis. Compared to a
classic high-pass filter with a smoothly varying response function, 
this method has the advantage that it removes only a small amount of power
and only in a very limited and well-defined frequency range, which was
important because of the expected low frequencies of the oscillations in
the stars.  %(evt FIG before and after for LaS star 17 maybe) 
%However, our approach still implies that any quantitative analysis of the
%stellar excess power and the search for regularly spaced peaks cannot be done
%for stars oscillating in the frequency range 0--50$\,\mu$Hz, which includes
%the clump stars (Nos. 21, 8, 9, 10, 20, 2 and 18; see Fig.~\ref{fig0} and
%Table~\ref{tab2}). 
The mean noise levels in the Fourier spectra at 300--900$\,\mu$Hz were
reduced by 2--13\%  (in amplitude) as a result of this cleaning process.
We did not decorrelate the time series against external parameters
(e.g. airmass, sky background, position on CCD) because we found that to    
have too dramatic and uncontrolled effects on the time series on time
scales similar to the expected stellar oscillations. %evt plots af tests 
This was based on decorrelation of simulated time series.

\begin{figure*}
 \includegraphics[width=160mm]{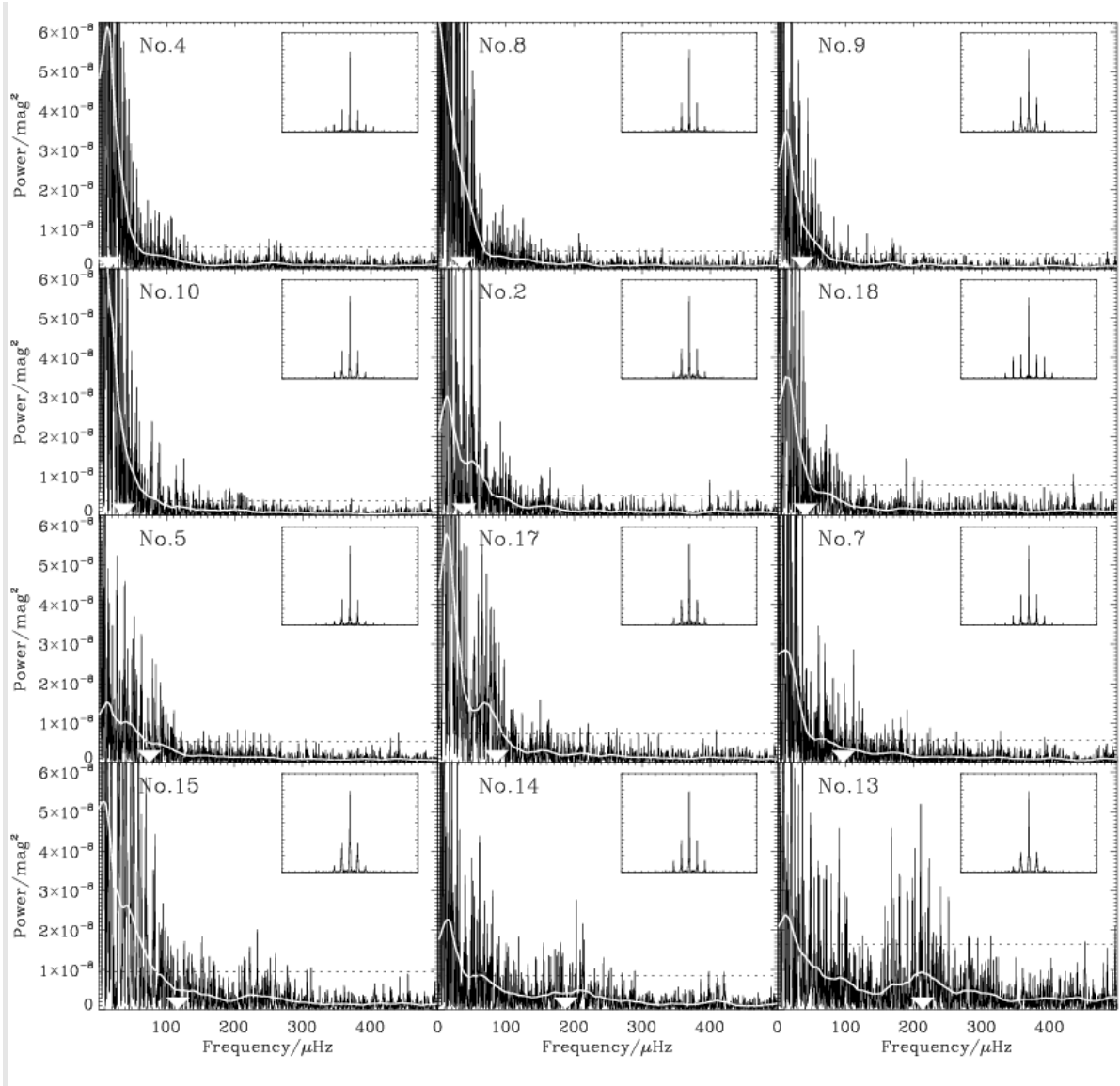}
 \caption{Fourier spectra of red giant stars (identifier shown in each
  panel). The solid white line is the smoothed spectrum, and the white
  arrow head indicates where we expect the stellar excess power
  (Table~\ref{tab2}). The dotted horizontal line shows
  $(3\sigma_{300\rmn{-}900\,\mu\rmn{Hz}})^2$. The inset shows the spectral
  window on the same frequency scale as the main panel.}   
\label{fig1}
\end{figure*}
After the initial cleaning of the dominant low frequency noise peaks, we
calculated the Fourier spectra of the 
combined data to search for excess power. 
The Fourier spectra were divided by the response function of the
cleaning process to restore the overall power distribution. The response
function was obtained by performing the same cleaning process on simulated
time series of white noise with the same window function as the
observations. The final response function for each star and each site was
the average Fourier spectrum based on 100 simulations.  
In Fig.~\ref{fig1} we show 
the spectra of the best twelve stars (noise below 50$\,\mu$mag in the
frequency range 300--900$\,\mu$Hz). Three other stars (Nos. 16, 19 and 20)
fulfill this criterion but have been omitted due to significantly higher
noise levels compared to stars of similar brightness. The solid white line
in each panel of Fig.~\ref{fig1} is the
smoothed spectrum, which was obtained by smoothing twice, with a
boxcar width of 30$\,\mu$Hz followed by one of 10$\,\mu$Hz. The frequencies
where we expect the stellar oscillations ($\nu_{\rmn{max}}$,
Eq.~\ref{eq_freq_max}) are indicated with the white 
downward-pointing arrow head, and the dotted horizontal line shows
$(3\sigma_{300\rmn{-}900\,\mu\rmn{Hz}})^2$. 
%We note that the detection threshold for stochastically excited and
%damped oscillations in a frequency region where the noise is not white
%depends on both mode lifetime and the characteristics of the noise source.
%Both are unknown in the present case, making a quantitative statement about
%upper limits of the oscillation amplitudes difficult. 
The increase in noise towards low frequencies, are due to drifts in the
data from instrumental and atmospheric instabilities as discussed in Paper
1. 

%-----------------------------------------------------------------------
\subsection{Location of excess power}
\begin{figure}
 \includegraphics[width=84mm]{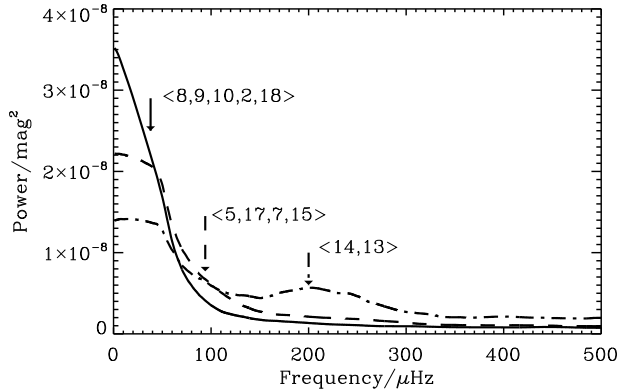}
 \caption{Average power distribution as a function of frequency for three
  groups of stars. The arrows show the mean position of the expected
  excess power, $\langle\nu_{\mathrm{max}}\rangle$, for the stars (stellar
  IDs in brackets) that have been used to calculate the corresponding
  power distribution  (see text for details).}   
\label{fig2}
\end{figure}

We expect the oscillation modes to be located at lower frequencies for the
more luminous stars (see Eq.~\ref{eq_freq_max} and Table~\ref{tab2}).
To search for general trends in the Fourier spectra, we grouped the stars
according to luminosity.  We formed three groups: the 
clump stars (5 stars), the stars between the clump and the lower RGB (4
stars), and the lower RGB stars (2 stars), but we excluded star No. 4 due
to its sole location in the colour-magnitude diagram (see Fig.~\ref{fig0}). 
For each group we averaged their Fourier spectra and smoothed the final
spectrum. Smoothing was done twice. First, with a wide boxcar
(width=100$\,\mu$Hz) to smear out humps of power originating from the 
individual spectra to better illustrate the overall distribution of power
within each group. We then used a second boxcar (width=10$\,\mu$Hz) to
smooth small point-to-point variations in the final plot. The result, 
shown in Fig.~\ref{fig2}, confirms the general trend of shifting power
excess as a function of luminosity, in agreement with expectations (see
arrows in Fig.~\ref{fig2} showing expected location of excess power). 
%in caption say arrows show the average position of the expected steller
%power within each group.

We investigated whether the observed power shift could have been 
produced by the reduction methods.
(i) We inspected Fourier spectra and weights from each
site and concluded that the power shift was not an artifact originating
from high weights given to a few sites. 
(ii) We reanalysed the data without removing the peaks at 11.57, 23.15,
34.72 and  46.30$\,\mu$Hz, but saw the same trend, although with more power
at low frequencies.
(iii) Finally, we investigated whether our data reduction could introduce
such a trend due to more pronounced drift noise for brighter and cooler
stars. 
%Could this simply be due to a larger aperture for the brigher stars during
%the extraction of the photometry?
The best comparison stars for this purpose were the blue stragglers, which
had similar brightness to the red giants and hence similar input parameters
for the extraction of the photometry 
\citep[for details see][~and Paper I]{KjeldsenFrandsen92}. 
However, their colours were very different from the red giants and from star
to star, and only a few blue stragglers were observed at most sites.
The blue stragglers showed different levels of excess power at low
frequencies from star to star, but no causal relation could be found.

In summary, we found no evidence for a non-stellar source that could
explain the observed pattern of excess power. The observed behaviour is
exactly as expected for oscillations. However, due to the lack of a
good ensemble of reference stars, similar to the red giant sample, we could   
not exclude with certainty that the photometric data reduction might have
caused the shift as a result of more drift noise for more luminous and
cooler stars.

%-----------------------------------------------------------------------
\subsection{Amplitude of excess power}
To search for further evidence of solar-like oscillations, we measured the
amount of excess power above the background noise, and converted that into
amplitude per mode in order to compare it with expected amplitudes.

\begin{figure*}
 \includegraphics[width=170mm]{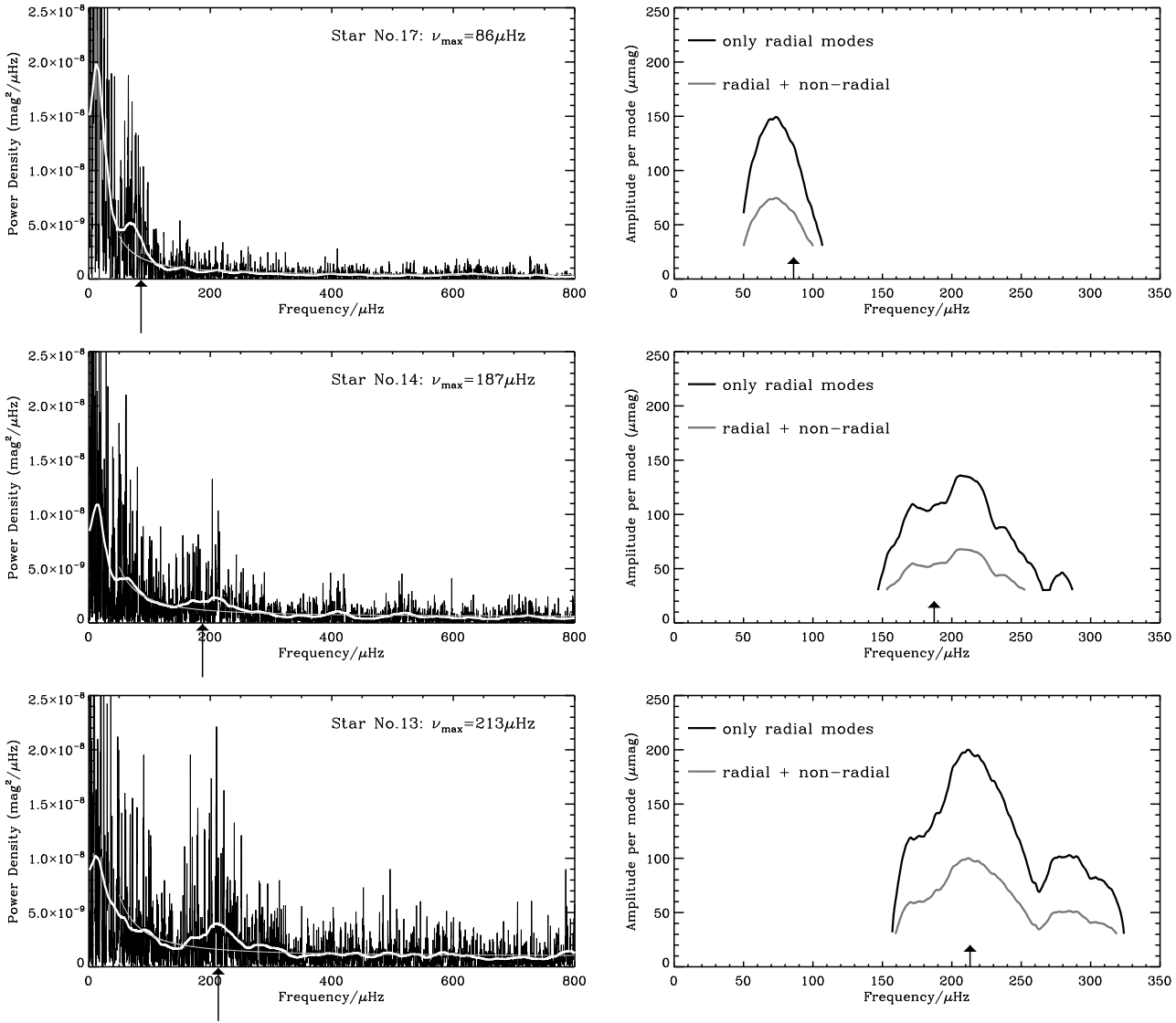}
 \caption{{\bf Left panels:} Power density spectra of three selected red
 giant stars. The thick white solid line is the smoothed spectrum and the
 thin grey line is a fit to the noise. The arrow indicates the
 expected value of
 $\nu_{\rmn{max}}$. Power density was calculated from Fig.~\ref{fig1} by
 dividing with the integrated power of the spectral window.    
 {\bf Right panels:} Estimated amplitudes for each of the three stars, 
 derived by subtracting the noise from the smoothed power density
 spectrum, multiplying with $\Delta\nu_{0}/n$ ($n=1$ for radial modes only,
 $n=4$ for radial plus non-radial modes) and taking the square root.} 
\label{fig3}
\end{figure*}

The noise level in the power spectrum, as a function of
frequency, $p(\nu)$, was estimated using the following expression:
%\mbox{$\sigma(\nu)=\frac{a}{\nu^{1.5}}+\sigma_{\rmn{wn}}$} (in
\mbox{$p(\nu)=a/\nu^{1.5}+p_{\rmn{wn}}$}. We tried $\nu$ to the power of
$-1$, $-1.5$ and $-2$, and found $-1.5$ provided the best fit in the region
around the excess power. We  measured the
white noise, $p_{\rmn{wn}}$, as the 
mean level in the frequency range 300--900$\,\mu$Hz. Then, having fixed
$p_{\rmn{wn}}$, we calculated $a$ requiring that the integral
of $p(\nu)$ in the range 50--900$\,\mu$Hz should equal that of the
power spectrum.
Assuming the excess power above this noise fit is real, we
estimated the amplitude per mode based on the approach of \citet{Kjeldsen05}.
These amplitude estimates are independent of the mode lifetime, but require
an assumption of the number of modes that are excited. We have estimated
the amplitudes for two extreme scenarios: (1) only radial modes are
excited and, 
(2) three additional non-radial modes per radial mode,
neglecting any difference in mode visibilities due to varying cancellations
in full-disk observations.  
%forgaaende saetning lav om saa lyder bedre 
The approach is illustrated in Fig.~\ref{fig3} for three stars that all
show a hump of power close to the expected frequency for solar-like
oscillations (see arrow in left panels). 
The power spectra (Fig.~\ref{fig1}) were divided by the integrated power of
the spectral window to obtain power density spectra (Fig.~\ref{fig3} left
panels). The background noise was subtracted from the smoothed spectra and
the resulting power density was multiplied by the mean 
mode spacing ($\Delta\nu_{0}$ for only radial modes, and $\Delta\nu_{0}/4$ for
both radial and non-radial modes). We then took the square root to convert
to amplitude (Fig.~\ref{fig3} right panels). 

\begin{figure}
 \includegraphics[width=84mm]{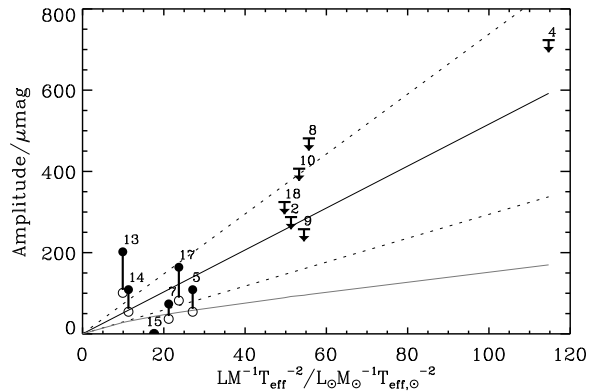}
 \caption{Measured amplitude versus $L/(M T_{\mathrm{eff}}^2)$ for the best 
 target stars (stellar IDs are indicated). Arrows show 
 observed upper limits for stars where we could not 
 estimate the noise contribution. For the other stars the filled dots
 (assuming only radial modes) and
 empty circles (assuming additional three non-radial modes) bracket
 the inferred amplitudes. The solid black line shows the 
 mean amplitude from 100 simulations using Eq.~\ref{loverm} as input
 amplitude. The dotted lines are $\pm 2\sigma$, where $\sigma$ is the
 rms scatter in the simulations. The grey line indicates the mean amplitude
 based on $(L/M)^{0.7}$-scaling}   
\label{fig2a}
\end{figure}

The amplitude at $\nu_{\mathrm{max}}$ for each star is plotted versus $L/(M
T_{\mathrm{eff}}^2)$ in Fig.~\ref{fig2a} and compared with results from
simulations which used input amplitudes derived from Eq.~\ref{loverm}. The
filled dots corresponds to the values read from the solid 
black curves in Fig.~\ref{fig3} (right panels), while the empty circles
in Fig.~\ref{fig2a} corresponds to the solid grey curves (Fig.~\ref{fig3}).
However,  for stars expected to oscillate at frequencies lower than
$\sim50\,\mu$Hz, we could not make realistic estimates of the noise
contribution due to the apparent rapid rise in noise at low frequency,
making it impossible to disentangle noise and stellar signal. We could only
obtain upper limits, which are 
indicated with arrows in Fig.~\ref{fig2a}. Since these are upper limits, we
only show the value derived assuming only radial modes are excited in these
stars.


Calculations by \citet{Dalsgaard04} and \citet{Dziembowski01} on more
massive and luminous stars than our targets suggest that only radial modes
are excited to observable amplitudes in red giant stars. No similar
theoretical pulsation analysis have been published for the red giants
in M67 in order to determine whether non-radial modes are expected.
However, stars Nos. 13 and 14 are expected to oscillate at frequencies that
are significantly higher than for typical red giant stars, and are similar
to those of subgiants. Observations of the subgiants $\eta\,$Boo
\citep{Kjeldsen03a}, $\beta\,$Hyi \citep{Bedding07} and $\nu\,$Ind
\citep{Carrier07} clearly show evidence of both radial and non-radial
modes.    
%{\bf For star No 13 the presence of non-radial modes are required to make the
%measured power excess in agreement with the scaling relation in
%Eq.~\ref{loverm}. }

The results in Fig.~\ref{fig2a} show that most stars are in agreement
with the scaling relation in Eq.~\ref{loverm} (illustrated with the black
line). There is a tendency that the scaling based on $(L/M)^{0.7}$ (grey
line) is in less good agreement with the data. However, we are not able to
exclude either due to the uncertainty whether non-radial modes are present,
and further because we only have upper limits for the most luminous stars.
We note that for star No. 14 some non-radial modes are required to make
the measured power excess in agreement (within $2\sigma$) with the scaling
relation in Eq.~\ref{loverm}. We further note that star No. 13 fall above
the $2\sigma$ region even in the extreme case of three additional
non-radial modes per radial mode. Hence, if the observed hump of power for
this star is due to 
stellar oscillations, it is likely that non-radial modes are excited, and
that we observed the star in a rare state of high amplitudes. The reason
for this high excess power could also be due to more
complicated drift noise in the data than described by our noise fit.

%For star No. 17, the measured amplitude is in good agreement with 
%$L/M$-scaling if we assume only radial modes are excited, while additional
%non-radial modes are required to match the prediction from
%$(L/M)^{0.7}$-scaling (see Table~\ref{tab4} and Fig.~\ref{fig3}, top right
%panel). However, the fit to the noise is quite uncertain at 
%these low frequencies. The hump of power, although
%more pronounced after cleaning the four dominant low
%frequency peaks, was also present before cleaning and so cannot
%be explained by simple drift noise.
%
%Star No. 14 shows excess power that is consistent with $L/M$-scaling only if
%both radial and non-radial modes are excited.
%The excess power seen in star No. 13 
%seems to be four times higher than predicted by $L/M$-scaling. Even if
%non-radial modes are excited we see twice the amount of predicted power.
%We note that the noise might not be well represented by our fit
%(Fig.~\ref{fig3}, right panels), and excess  
%power relative to this fit could be due to more complicated drift noise
%in the data.

%The measured amplitudes of stars Nos. 17, 14 and 13 using this approach are in
%agreement with what we find if we try to match the observed Fourier spectra
%with simulations. 

In summary, we see good evidence for oscillation power in three out of six
stars that are expected to oscillate with frequencies larger than
$50\,\mu$Hz. For stars expected to oscillated below $50\,\mu$Hz, we were
only able to provided upper limits due to difficulties in determining the
noise levels in their power spectra. In general we see agreement
with the predicted amplitudes from $L/M$-scaling. 

\subsection{Autocorrelation of excess power}
\begin{figure}
 \includegraphics[width=84mm]{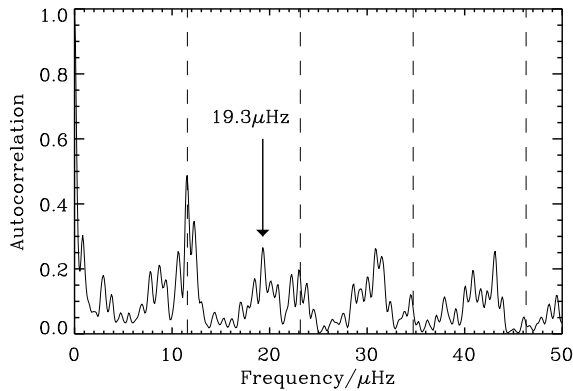}
 \caption{Autocorrelation of star No. 13 (see text). Dashed lines indicate
 integer multiples of one cycle per day (11.574$\,\mu$Hz).}  
\label{fig6}
\end{figure}

To search for a regular series of peaks, which is a typical signature of
solar-like oscillations, we calculated the autocorrelation of all the stars
in the frequency range where we expect stellar oscillations.  
For most stars the autocorrelation is dominated by the aliases at 1c/d. In
about half the stars there is a hint of a peak that coincides with 
the expected large frequency separation (Eq.~\ref{eq_large_sep}) and of
those, there are two (Nos. 7 
and 13) that show intriguing results. In the former, the most prominent
peak is located at
8.8$\,\mu$Hz, which is very close to the expected value (Table~\ref{tab2}).
%Star No. 17 showed only peaks in
%the autocorrelation at 1/2 and 1 cycle per day, and every
%combination of these. The autocorrelation of star No. 14 reveals no clear
%frequency separation, showing only peaks corresponding to the daily
%aliases. 
In Fig.~\ref{fig6} we show the autocorrelation of
star No. 13. The autocorrelation was calculated in the
frequency range 135--285$\,\mu$Hz setting all values in the Fourier
spectrum lower than a threshold of $2\sigma_{300\rmn{-}900\,\mu\rmn{Hz}}$
(in amplitude) equal to the threshold value. 
The most significant peak that is not a daily alias is located at
19.3$\,\mu$Hz, which is within 15\% of the expected value
(Table~\ref{tab2}).

%Hya did not show clear sep either depite the much higher S/N
%but is still believed to be stellar oscillations, so the lack of an obvious
%spacing in this low S/N case does not exclude stellar origin of excess
%power.

%-----------------------------------------------------------------------
\subsection{Stellar granulation}\label{granulation}
In order to estimate the signal from background granulation we use a
simple scaling of the solar granulation background observed by SoHO (VIRGO
green channel). The scaling uses the Harvey model \citep{Harvey85} to
describe the size and 
shape of the granulation power spectrum. The Harvey model needs as input the
characteristic time scale for granulation, as well as the granulation scatter
in the time series. We calibrated those two parameters using the solar data,
using the following simple assumptions:

(1) The size of a granulation cell on the surface of a star is proportional
to the scale height of the atmosphere at the stellar surface. The variance of
the granulation in the time series is then assumed to be inverse
proportional to the total number of cells on the surface.

(2) The characteristic time scale of granulation is approximated to be
proportional to the ratio between the size of the individual cells and the
velocity of the cell. We estimate the granulation cell velocities to be
proportional to the sound speed in the stellar atmosphere.

(3) We assume that the atmosphere is isothermal (allowing a simple scaling
of the atmospheric scale height).

\begin{figure*}
 \includegraphics[width=160mm]{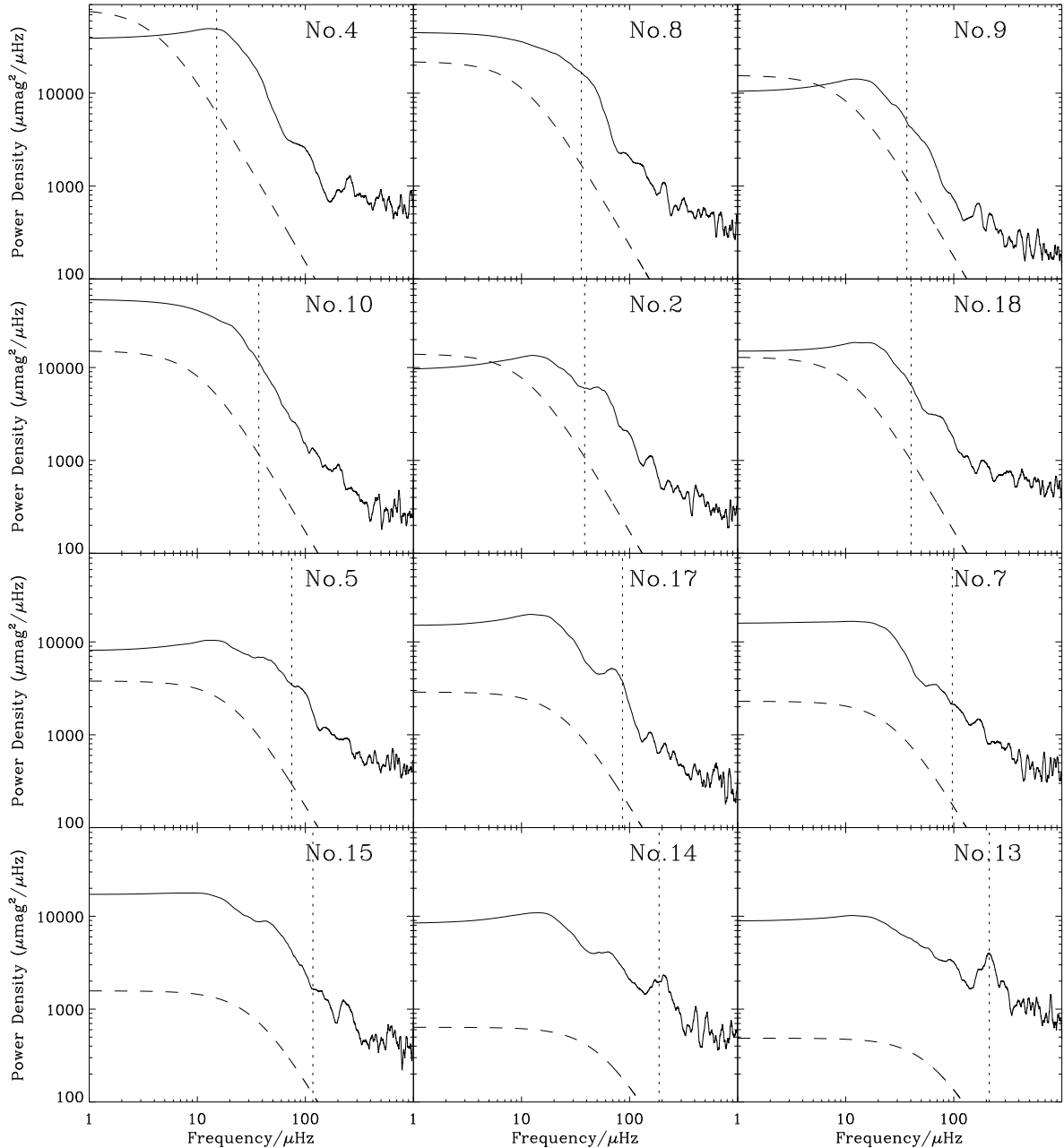}
 \caption{Smoothed power density spectra of red giant stars (identifier
  shown in each panel). The dashed curve is the estimated granulation signal
  based on scaling from the Sun (see Sect~\ref{granulation}). The dotted
  line indicates where stellar oscillations are expected
  (Eq.~\ref{eq_freq_max}).} 
\label{fig2b}
\end{figure*}

%Those assumptions and the Harvey model predict that there is a peak power in
%the granulation power spectrum if we plot power density times frequency
%squared as a function of the frequency. The peak will appear at the
%characteristic frequency.
More details can be found in Kjeldsen \& Bedding (in preparation).
Using these assumptions, we are able to calculate granulation power spectra
that agrees both in shape and level with detailed simulations by
\citet{SvenssonLudwig04}.
In Fig.~\ref{fig2b} we plot smoothed power density spectra of the target
stars, together with the estimated signal from background granulation. For
the most luminous stars granulation is likely to contribute significantly at
low frequencies, while for the most faint stars
in our sample granulation can be excluded as a possible cause of
increasing noise.
From this figure we can also see that there is a distinct difference in
time scale between the granulation and the humps of power that appear to be
due to stellar oscillations. Background granulation will therefore not prevent
detection of solar-like oscillations.

%-----------------------------------------------------------------------
\section{Simulations}\label{simulations}
\begin{figure}
 \includegraphics[width=84mm]{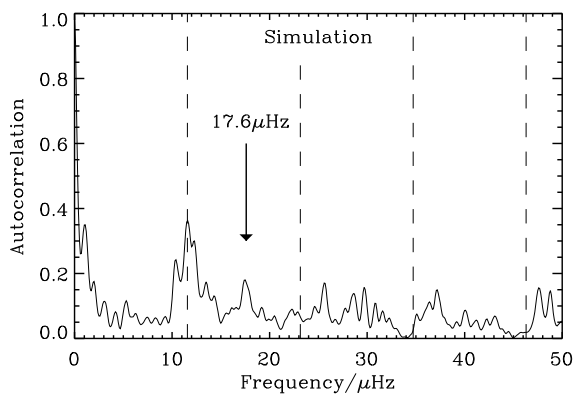}
 \caption{Autocorrelation of a simulation of star No. 13 (see text). Dashed
 lines indicate integer multiples of one cycle per day (11.574$\,\mu$Hz).}  
\label{fig7}
\end{figure}
To interpret the results shown in Sect.~\ref{analysis}, we made a series of
simulations of each target star using the method described by
\citet{Stello04}. We chose a regular series of input frequencies with
a separation of $\Delta\nu_{0}$ (Table~\ref{tab2}). Their relative
amplitudes were determined by a Gaussian envelope with a height
corresponding to the $L/M$-scaling values in Table~\ref{tab2}
(note that $(L/M)^{0.7}$-scaling predicts amplitudes that are roughly 0.3--0.5
times that for our targets). 
The envelope was centered at $\nu_{\rmn{max}}$ (Table~\ref{tab2}) with a
width equal to $0.48\,\nu_{\rmn{max}}$, which was calibrated to the
observations of $\xi\,$Hya \citep{Stello04}. The envelope reproduced by
this approach is in good agreement with $\beta\,$Hyi and the Sun 
\citep{Kjeldsen05}. 
The number of input frequencies was the closest integer of
$7\times0.48\,\nu_{\rmn{max}}/\Delta\nu_{0}$, which ensures that we
have frequencies within the entire envelope. Because the non-white noise
is difficult to estimate precisely, we chose to include only white noise in  
the simulations, which we 
set equal to the mean level in the Fourier spectra at 300--900$\,\mu$Hz
of the observed data.  
For each target star we ran simulations in two parallel runs, one with a 
mode lifetime of 20 days, in agreement with the theoretical calculations on
the red giant $\xi\,$Hya \citep{HoudekGough02}, and the other of 2 days
in agreements with the observations of that star \citep{Stello06}.

\begin{figure*}
 \includegraphics[width=160mm]{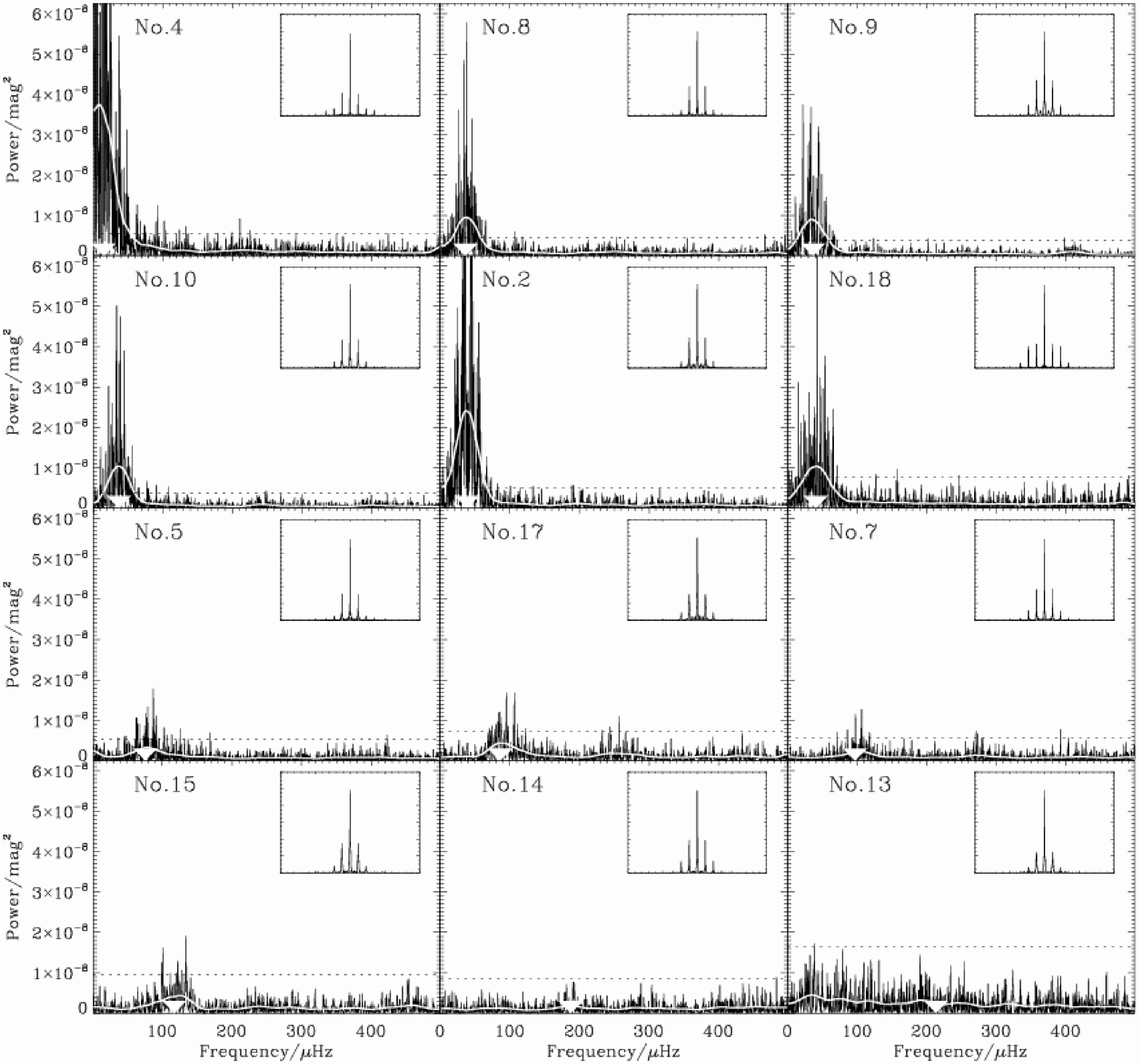}
 \caption{Fourier spectra of simulated red giant stars (identifier shown in
  each panel), with a mode lifetime of 2 days and amplitudes according to
  $L/M$-scaling (Eq.~\ref{loverm}). The solid white line is the smoothed
  spectrum, and the white arrow head indicates where we expect the stellar
  excess power (Table~\ref{tab2}). The inset shows the spectral window on
  the same frequency scale as the main panel. The dotted line is
  $(3\sigma_{300\rmn{-}900\,\mu\rmn{Hz}})^2$.}   
\label{fig4}
\end{figure*}

\begin{figure}
 \includegraphics[width=84mm]{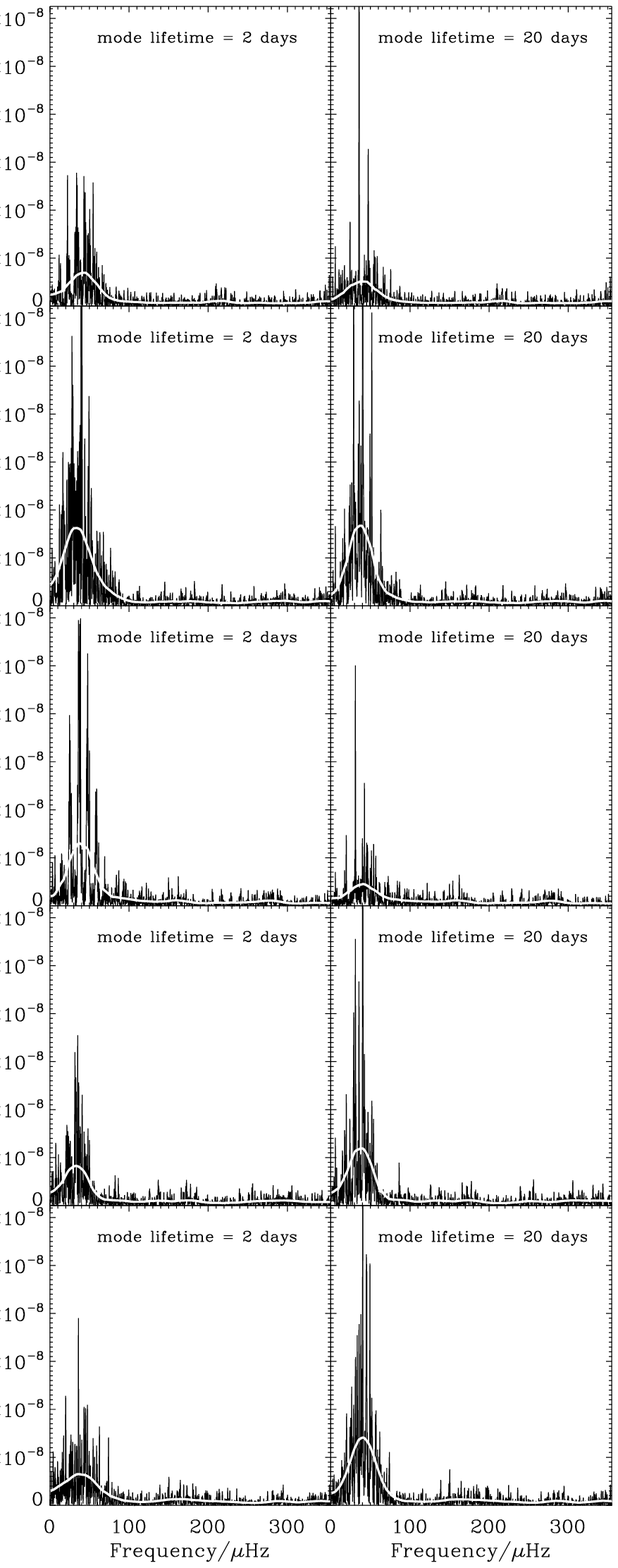}
 \caption{Fourier spectra of simulated data (star No. 2) for five different
  random number seeds and two different mode lifetimes (left: 2 days,
  right: 20 days). The solid white line is the smoothed spectrum.}  
\label{fig5}
\end{figure}

We show here the results for the short mode lifetime (Fig.~\ref{fig4}). The
difference between the long and short mode lifetime is discussed below.
It is important to
stress that the detailed characteristics of the excess power in the Fourier 
spectra depend very much on the complicated interaction between the
spectral window and the oscillation modes (their frequencies, amplitudes
and lifetime), as well as the random number seed. For example, using the
same random number seed, the clump stars (Nos. 8, 9, 10, 2 and 18), which had
different spectral windows but only slightly different amplitudes and input 
frequencies, show very different power excesses (Fig.~\ref{fig4}).
For star No. 2 we show in Fig.~\ref{fig5} (left panels) five spectra based
on different random number seeds. Again, we see a large variation in the
result, which is in agreement with Fig.~\ref{fig2a} (dashed lines). 
This large intrinsic variation was also pointed out by
\citet{Stello04,Stello06}.  The right panels show the  
corresponding spectra based on the long mode lifetime (=$20\,$days). From
this, we see that a long mode lifetime in general provides higher peaks,
but the 
differences are comparable to the difference that arises from using a
different seed.

A comparison of observations and simulations (Figs.~\ref{fig1} and \ref{fig4})
indicates that a significant part of the excess power seen in the
observations might be of stellar origin. The excess power in the
observations can be associated with either humps of power (see Nos. 13 and
17 in Fig.~\ref{fig1}) or a broad ``shoulder'' on top of the drift noise
(see No. 2). It is 
clear from the simulations that observing excess power in one star does not
imply that we necessarily would see a similar hump in a similar star (or
even in the same star if observed at another epoch; Fig.~\ref{fig5}).
We note that the simulations shown in Fig.~\ref{fig4} only included
radial modes. If the stars have non-radial modes these simulations
underestimate the stellar excess power. This could be significant at least
for the less evolved stars (Nos. 13 and 14). 
Very recently, \citet{Hekker06a} reported that non-radial modes
are present in two red giant stars ($\varepsilon\,$Oph and
$\eta\,$Ser) that oscillate at approximately
$60\,\mu$Hz  and  $130\,\mu$Hz, respectively, and hence are comparable to
stars in our sample. 
Here we also note that the scaling relation that was used to estimate the
amplitudes is quite uncertain for red giant stars. 

Finally, we calculated the autocorrelation of simulated data of star No. 13
that reproduced the observed excess power. This was done by setting the
input amplitude of the simulations to $200\,\mu$mag and including only
radial modes in accordance with Fig.~\ref{fig3}. This time we set the white
noise equal to the value estimated 
at 200$\,\mu$Hz from our noise fit in Fig.~\ref{fig3}. 
The autocorrelation was based on simulations with a short mode lifetime
(= 2 days) because they produced Fourier spectra more like the observed
spectrum.
In total nine independent simulations were made. In roughly half the 
cases we saw a peak in the autocorrelation of the same significance as
in the observations (Fig.~\ref{fig6}). In Fig.~\ref{fig7} we show one of
the examples which show a peak. However, these peaks were not at exactly
the same frequency but appeared in an interval from approximately 
16$\,\mu$Hz to 19$\,\mu$Hz. Some of these (as in Fig.~\ref{fig7}) were
close to halfway between 1 and 2 cycles per day
(17.4$\,\mu$Hz), and might therefore be caused by aliasing. In the
other cases the only clear peaks were those clearly originating from the
daily aliases at 1, 2 and 3 cycles per day.  

To investigate whether the peak seen at 19.3$\,\mu$Hz in the
autocorrelation of the observations was caused by interaction
between the spectral window and random peaks we repeated the simulations
using randomly spaced frequencies. For each of the three sets 
of random frequencies we made nine independent simulations, and inspected
the autocorrelation. The differences seen between each of the independent
simulations within each set of frequencies were small compared to the
difference between different sets. 
In most cases we did not see a clear peak in the range 16--19$\,\mu$Hz, but
in a few cases a clear peak halfway between 1 and 2 cycles per day
(17.4$\,\mu$Hz) was seen. In some cases peaks as significant as the
observed 19.3$\,\mu$Hz peak were seen at other frequencies.  
Hence, we conclude that the peak at 19.3$\,\mu$Hz in the observations could
be caused by randomly distributed peaks.

%\section{Rodekassen}
%We note that \citet{Gilliland93} used coherent oscillations in there
%simulations. Their upper limits are therefore more optimictic than if
%damping was taken into acount.

%Even if we had not been plagued by drift noise the expected narrow envelope
%of power excess with only very few modes excited makes it difficult to
%establish the solar-like nature of any excess power at low frequency (this
%can be tested in my sim...do that before saying this).

%-----------------------------------------------------------------------
\section{Conclusions}\label{conclusions}
We have analysed high-precision photometric time series of 20 red
giant stars in the open cluster M67. The data, recently published in
Paper I, were from a large multisite campaign. %, which provided the lowest
%noise and best coverage of and ensemble of red giant star. 
%The final characteristics of the non-white noise including drift in this
%complex data set could not be well/realistically estimated prior to
%observation. 

We see evidence of excess power in the Fourier spectra,
shifting to lower frequencies for more luminous stars, consistent with
expectations from solar-like oscillations (see Fig.~\ref{fig2}). %We note
				%that the method used to 
				%extract the photometric time series could  
%introduce more drift noise (hence more power at low frequencies) for the
%more luminous and cooler stars, which will simulate a shift of excess power.

We further estimated the mode amplitude from the observed power
excess assuming that it originates from solar-like oscillations. These
estimated amplitudes are in agreement with the results from simulations
using estimated amplitudes by scaling from the Sun (Fig.~\ref{fig2a}). 
These results further support that we have seen evidence for solar-like
oscillations in some of the red giant stars in M67. 
We note that the measured amplitudes from our simulations show a 
scatter of roughly 40\% from one epoch to the next, which is similar to
what we see in the Sun \citep{Kjeldsen07}. 
This intrinsic scatter should always be taken into account when evaluating 
observed amplitudes of solar-like oscillations. 
Without firm knowledge of whether non-radial modes are present in these
stars and with only upper limits on the amplitudes of the most luminous
stars, the large scatter in the measured amplitude limits our ability to
clearly pinpoint a favoured scaling relation for amplitude. However, we see
an indication of a better match between observations and the $L/M$-scaling 
\citep{KjeldsenBedding95} than with $(L/M)^{0.7}$-scaling
\citep{Samadi05}. 

In many stars we see apparently high levels of non-white noise, but its
temporal variation is unknown and could not be decorrelated. In the Fourier
spectrum this noise 
seems to be significant at frequencies below 100$\,\mu$Hz. This is
unfortunate because the target stars in which we expected the most clear
detections are expected to be oscillating in the affected frequency range.
We are therefore not able to disentangle the noise and stellar signal in
the analysis of these stars. Hence, the data did not allow the clear
detections that we had anticipated. 
Simple scaling of the solar granulation \citep{Harvey85} shows that the
power at low frequencies (lower than the expected p-mode range) is not
coming from granulation for the fainter 
stars on the low RGB, and is most likely instrumental (see
Fig.~\ref{fig2b}). However, it could be granulation 
noise in the more luminous clump stars and star No. 4 (see Fig.~\ref{fig0}).
%Due to the difficulties in estimating the noise level at low frequencies we
%do not find basis for extracting individual frequencies. 
%Theoretical modelling of the pulsations in these stars would greatly
%improve our capability to interpret our results.  

%In the targets were we do not see any
%evidence of excess power upper limits of the amplitudes are not given. This
%is because the noise level in the relavant frequency range is unknown.

If the observed power excesses are due to stellar oscillations this result
shows great prospects for asteroseismology on clusters. However, with the
unfortunate cancellation of the Eddington mission by ESA, we might have to
wait many years for a dedicated space project specifically aiming at
asteroseismology on stellar clusters. 
Alternatively, one could initiate a ground-based network with
high-resolution and high-sensitivity spectrographs that is able to detect
oscillations in faint cluster stars, or a photometric multisite campaign
with larger telescopes (preferably larger than 2 meter, see Paper I)
with stable instrumentation located at good sites aimed at red giants in
clusters. As clearly demonstrated by our results, the final level of the
non-white noise 
(including drift) in the data will be absolutely crucial in any approach to
detect solar-like oscillations in red giant stars. This largely favours space
missions as atmospheric effects in ground-based data varies on time scales
similar to the stellar oscillations.
Although the Kepler and COROT space missions will not specifically target
clusters it is expected that these missions will observe many red giants,
some of which would be in clusters. Recent results from the  
MOST satellite on the red giant $\varepsilon\,$Oph \citep{Barban07}
demonstrates that space missions have the potential to 
do very interesting science on red giant field stars, and more K giants are
on the target list for MOST in the future.

We would be happy to make the data presented in this paper available on
request.

\section*{Acknowledgments}
This research was supported by the Danish Natural Science Research Council
through its centre for Ground-Based Observational Astronomy, IJAF, and by
the Australian Research Council.
This work was partly supported by the Research Foundation Flanders (FWO).
This paper uses observations made from the South African Astronomical
Observatory (SAAO), Siding Spring Observatory (SSO), Mount Laguna
Observatory (MLO), San Diego State University, and the Danish 1.5m
telescope at ESO, La Silla, Chile. We will be happy to provide data on
request.

\bibliography{bib_complete}

\end{document}